\documentclass[twocolumn,showpacs,preprintnumbers,amsmath,amssymb]{revtex4}

\usepackage{graphicx}
\usepackage{dcolumn}
\usepackage{bm}
\def\jpsi{{J/\psi}}
\def\hats{{\hat{s}}}
\def\li{{\mathrm{Li_2}}}

\def\be{\begin{equation}}
\def\ee{\end{equation}}
\def\bea{\begin{eqnarray}}
\def\eea{\end{eqnarray}}
\def\NO{\nonumber}
\def\gev{\mathrm{~GeV}}

\def\fb{\mathrm{~fb}}
\def\pb{\mathrm{~pb}}

\def\dfrac{\displaystyle\frac}
\def\co{{\cal O}}
\def\a{\alpha}
\def\b{\beta}

\def\d{\delta}
\def\e{\epsilon}

\def\s{\sigma}

\begin{document}


\title{Next-to-Leading-Order QCD Corrections to $e^+e^-\rightarrow \jpsi g g$ at the B Factories }

\author{Bin Gong and Jian-Xiong Wang}%
\affiliation{
Institute of High Energy Physics, CAS, P.O. Box 918(4), Beijing, 100049, China. \\
Theoretical Physics Center for Science Facilities, CAS, Beijing, 100049, China.
}%
\date{\today}

\begin{abstract}{\label{abstract}}
We calculate the next-to-leading-order (NLO) QCD corrections to
$e^+e^-\rightarrow \jpsi g g$ via color singlet
$\jpsi(^3\hspace{-0.8mm}S_1)$ at the B factories. The result shows
that the cross section is enhanced to $0.373\pb$ by a K factor
(NLO/LO) of about $1.21$. By considering its dependence on the charm
quark mass and renormalization scale, the NLO cross section can
range from $0.294$ to $0.409\pb$. Further including the $\psi^\prime$ feed-down,
$\sigma(e^+e^-\rightarrow \jpsi X(\mathrm{non}\ c\bar{c}))$ is
enhanced by another factor of about $1.29$ and reaches $0.482\pb$. In
addition, the momentum distributions of $J/\psi$ production and polarization 
are presented. Recent measurements from Belle agree well with our prediction for the cross section and
momentum distribution. It is expected that this process can serve as a very good channel to clarify
the $J/\psi$ polarization puzzle by performing further experimental measurements.
\end{abstract}

\maketitle {\label{introduction}} Perturbative quantum
chromodynamics are successful to describe large momentum transfer
processes at quark level due to its asymptotic freedom property. But
it falls into the nonperturbative range for quark hadronization to
form the final state hadrons which are measured in experiments.
Therefore the quark hadronization is usually described by
phenomenological models and quite far away from the first principle QCD
Theory. However, in the heavy quarkonium case, a naive perturbative QCD
and nonrelativistic treatment for the bound state is applied straightforwardly
to the related decay or production processes. It is
called color-singlet mechanism. To describe
the huge discrepancy of the high-$p_t$ $J/\psi$ production between
the theoretical calculation based on color-singlet mechanism and the
experimental measurement at Tevatron, a color-octet
mechanism~\cite{Braaten:1994vv} was proposed based on the
non-relativistic QCD (NRQCD)~\cite{Bodwin:1994jh}. It allows
consistent theoretical predictions to be made and to be improved
systematically in the QCD coupling constant $\a_s$ and the
heavy-quark relative velocity $v$. 
In recent years, there is a huge data collection in the B factory
experiments. Based on that, many $\jpsi$ production processes were
observed~\cite{Abe:2001za,Abe:2002rb,Aubert:2005tj} in the past. Now
the integrated luminosity is more than $850\fb^{-1}$ at the Belle
detector at the KEKB and it is about 20 times larger than the
integrated luminosity $32.4\fb^{-1}$, based on which the inclusive
$\jpsi$ production was measured ~\cite{Abe:2001za,Abe:2002rb}.
Therefore it supplies a very important chance to perform
systematical study on $\jpsi$ production both theoretically and
experimentally.

The measurements for exclusive $\jpsi$ productions $e^+e^-
\rightarrow \jpsi  \eta_c$, $\jpsi  \jpsi$, $\jpsi
\chi_{cJ}$ at the B factories have shown that
there are large discrepancies between the leading-order (LO)
theoretical
predictions~\cite{Braaten:2002fi,Liu:2002wq,Hagiwara:2003cw,Bodwin:2002fk}
in NRQCD and the experimental
measurements~\cite{Abe:2002rb,Aubert:2005tj,Abe:2004ww}. It seems
that such discrepancies can be resolved by introducing higher order
corrections~\cite{Braaten:2002fi,Zhang:2005ch,Gong:2007db,Gong:2008ce,Bodwin:2002kk}:
next-to-leading-order (NLO) QCD corrections and relativistic
corrections.

The cross section for inclusive $\jpsi$ production in $e^+e^-$
annihilation was measured by
BABAR~\cite{Aubert:2001pd,Aubert:2005tj} as $2.54\pm0.21\pm0.21\pb$
and Belle~\cite{Abe:2001za, Abe:2002rb} as $1.45\pm0.10\pm0.13\pb$.
These measurements include both $\jpsi+c\bar{c}+X$ and
$\jpsi+X(\mathrm{non}~c\bar{c})$ parts in the final states. Many
theoretical
studies~\cite{Driesen:1993us,Yuan:1996ep,Cho:1996cg,Schuler:1998az,Baek:1998yf,
Liu:2002wq,Hagiwara:2007bq, Braaten:1995ez,Wang:2003fw} have been
performed on this production at LO in NRQCD and the results for inclusive $J/\psi$
production cover the range $0.6\sim1.7\pb$ depending on parameter choices. A further
analysis by Belle~\cite{Abe:2002rb} gives \bea
\sigma(e^+e^-\rightarrow \jpsi c\bar{c}+X)=0.87^{+0.21}_{-0.19}\pm
0.17\pb. \eea It is about 5 times larger than the LO NRQCD
prediction~\cite{Liu:2002wq}. 
However, this large discrepancy was partially
resolved by considering both NLO correction and feed-down from higher
excited states \cite{Zhang:2006ay}. 
The above measurements infer
that $\sigma[e^+e^-\rightarrow \jpsi+X
(\mathrm{non}~c\bar{c})]=0.6\pb$. For this part, the contributions
from the color-singlet and color-octet contributions for the
processes, $e^+e^-\rightarrow
\jpsi^{(1)}(^3\hspace{-0.8mm}S_1)gg,\jpsi^{(8)}(^1\hspace{-0.8mm}S_0,
^3 \hspace{-0.8mm}P_J)g$, are about $0.2\pb$ and $0.27\pb$,
respectively, at the LO in NRQCD~\cite{Wang:2003fw}. However, the
signal of the color octet was not found in the
experiment~\cite{Aubert:2001pd,Abe:2001za}. Therefore, the
experimental measurement by Belle is about 3 times larger than the
theoretical prediction from color singlet at LO, and can be much
more than 3 times by BABAR.
\begin{figure*}
\center{
\includegraphics*[scale=0.58]{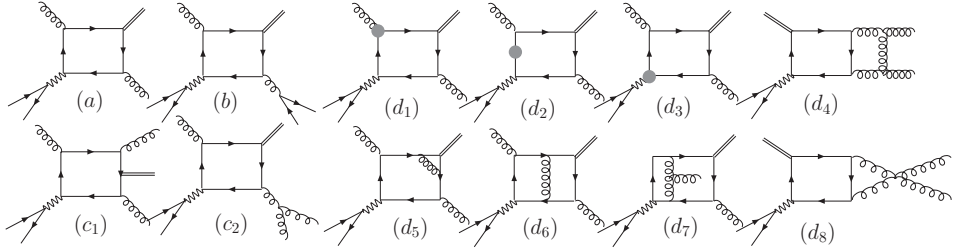}
\caption {\label{fig:Feynman}Typical Feynman diagrams. (a): LO; (b) $e^+e^-\rightarrow \jpsi  gq\overline{q}$;
(c) $e^+e^-\rightarrow \jpsi g g g$; (d): One-loop. Groups $(d_1)-(d_3)$ are the counter-term diagrams,
including corresponding loop diagrams.}}
\end{figure*}
Recently many
studies~\cite{Zhang:2005ch,Zhang:2006ay,Gong:2007db,Gong:2008ce,Gong:2008hk,Gong:2008ft,Gong:2008ue}
have shown that higher order corrections within NRQCD framework are
very important. To achieve a reasonable theoretical prediction for
the process $e^+e^-\rightarrow \jpsi g g$, in this letter, we
present a NLO QCD calculation to this process.

The related Feynman diagrams which contribute to the LO amplitude of the process $e^+(p_1) + e^-(p_2) \rightarrow \jpsi(p_3) + g(p_4) + g(p_5)$ are shown
in Fig.~\ref{fig:Feynman}$(a)$, while the others can be obtained by permuting the
places of the virtual photon and gluons.
In the nonrelativistic limit, the NRQCD factorization formalism is used to obtain
the total cross section in $n=4-2\e$ dimensions as
\begin{widetext}
\bea
\s^{(0)}&=&
-\dfrac{\a^2\a_s^2e_c^2|R_s(0)|^2}{9m_c^5\hats^3(\hats-1)}
\biggl\{18(\hats^2-2\hats+2)
+\dfrac{2\hats(5\hats^2-14\hats+3)}{\hats-1}\ln(\hats)
+\dfrac{2(4\hats^2-9\hats+8)\beta}{\hats-1}\ln(2\hats-1-2\beta) \NO\\ &&+\dfrac{2\hats^3-\hats^2-12\hats+8}{\hats-1} \left[\li\left(\dfrac{2(\hats-1)}{\hats+\beta-1}\right)
+\li\left(\dfrac{2(\hats-1)}{\hats-\beta-1}\right)\right]\biggr\}
+ \co(\e),
\label{eqn:CS_LO}
\eea
\end{widetext}
by introducing a dimensionless kinematic variable $\hats=s/(2m_c)^2$,
where $s$ is the squared center-of-mass energy, $e_c$ and $m_c$ are the electric charge and mass of the charm quark, respectively, and $\beta=\sqrt{\hat{s}(\hat{s}-1)}$. $R_s(0)$ is the radial wave function at the origin of $\jpsi$. The approximation $M_{\jpsi}=2m_c$ is taken.

At NLO in $\a_s$, there are virtual corrections which arise from
loop diagrams. Dimensional regularization has been adopted for
isolating the ultraviolet (UV) and infrared (IR) singularities.
UV-divergences from self-energy and triangle diagrams are canceled
upon the renormalization of QCD. 
Here we adopt the same renormalization scheme as Ref.~\cite{Gong:2007db}. There are
111 NLO diagrams in total, including counter-term diagrams. They are
shown in Fig.~\ref{fig:Feynman}$(d)$, and divided into 8 groups.
Diagrams of group $(d_5)$ contain Coulomb singularities, which can
be isolated
and mapped into the $c\bar{c}$ wave function.
Although the Feynman diagrams are similar, the calculation of tensor and scalar
integrals is much more complicated
than that in Ref.~\cite{Gong:2008hk}, because there is one more variable,
which is the mass of the virtual photon, in this calculation. Again, the calculation was done automatically
with our Feynman Diagram Calculation
package (FDC)\cite{FDC}.

The real corrections arise from two processes, $e^+e^-\rightarrow \jpsi gq\overline{q}$ and
$e^+e^-\rightarrow \jpsi g g g$. The related Feynman diagrams for these two processes are shown in
Fig.~\ref{fig:Feynman}$(b)$ and \ref{fig:Feynman}$(c)$.
The phase space integration for them will generate IR singularities,
which are either soft or collinear and can be conveniently isolated
by slicing the phase space into different regions. We adopt the two-cutoff
phase space slicing method \cite{Harris:2001sx} to decompose the phase space into three parts by introducing two small cutoffs, $\d_s$ and $\d_c$. And
then the real cross section can be written as
\be
\s^R=\s^S+\s^{HC}+\s^{H\overline{C}},
\ee
where $\s^S$ from the soft regions contains
soft singularities and is calculated analytically under soft approximation.
It is easy to find that the soft singularities for a gluon emitted from the
charm quark pair in the S-wave color singlet $\jpsi$ are canceled
by each other. And we have
\bea
d\s^S&=&d\s^{(0)} \frac{\a_s}{2\pi} \frac{\Gamma(1-\e)}{\Gamma(1-2\e)}
\left(\frac{4 \pi \mu^2}{s}\right)^{\e}
\left(\dfrac{A^S_2}{\e^2} + \dfrac{A^S_1}{\e} + A^S_0\right), \NO\\
A^S_2&=&6, \quad A^S_1= -12\ln\d_s
-6\ln\left(\sin^2\frac{\theta_{g}}{2}\right),
\NO\\
A^S_0&=&\dfrac{(A^{S}_1)^2}{12} +
6\li\left(\cos^2\frac{\theta_{g}}{2}\right), \label{eqn:soft} \eea
where $\mu$ is the renormalization scale and $\theta_{g}$ is the
angle between two gluons in the $p_1+p_2$ rest frame. $\s^{HC}$ from
the hard collinear regions contains collinear singularities and can
also be factorized. Here we have \bea d\s^{HC}&=&d\s^{(0)}
\frac{\a_s}{2\pi} \frac{\Gamma(1-\e)}{\Gamma(1-2\e)}
\left(\frac{4 \pi \mu^2}{s}\right)^{\e}\left(\dfrac{A_1^{HC}}{\e} + A^{HC}_0\right),\NO\\
A^{HC}_1&=&11+6\ln\d_s^{(4)}+6\ln\d_s^{(5)}-\frac{2}{3}n_{lf},\\
A^{HC}_0&=&\frac{67}{3}-\frac{10}{9}n_{lf}-2\pi^2 \NO\\
&&-3\ln^2\d_s^{(4)}-3\ln^2\d_s^{(5)} -\ln\d_cA^{HC}_1, \NO
\label{eqn:coll} \eea where $\d_s^{(j)}=\d_s/[1-(p_3+p_j)^2/s]$ and
$n_{lf}$ is the number of active light quark flavors. The hard
noncollinear part $\s^{H\overline{C}}$ is IR finite. Finally, all
the IR singularities are canceled analytically. After adding all the
contribution together, the cross section at NLO can be expressed as
\be
\s^{(1)}=\s^{(0)}\left\{1+\dfrac{\a_s(\mu)}{\pi}\left[a(\hats)+\b_0\ln\left(\dfrac{\mu}{2m_c}\right)\right]\right\},
\label{eqn:CS_NLO} \ee where $\b_0$ is the one-loop coefficient of
the QCD beta function.

To study the polarization of $\jpsi$ production, we define
the angular distribution $A$ and polarization factor $\alpha$ 
as:
\bea 
\dfrac{d^2\s}{d\cos\theta dp}=S(p)[1+A(p)\cos \theta],\quad
\alpha=\dfrac{\sigma_T-2\sigma_L}{\sigma_T+2\sigma_L},
\eea
where $p$ and $\theta$ are the 3-momentum and production angle of $\jpsi$ in the laboratory frame.
$\sigma_T$ and $\sigma_L$ 
are the transverse and longitudinal polarized cross section.
To calculate $\alpha$, we use the
same method to represent the polarized cross section as Eqs.~(8) and
(9) in Ref.~\cite{Gong:2008hk}. This method is found numerically
unstable in a small region of phase space due to the cancellation of
large numbers. Therefore, the momentum distributions for $A$ and
$\alpha$ contain potentially large numerical errors in our calculation for $p<0.5$ GeV or
$p>4.2$ GeV. As regards the total cross
section and momentum distribution of $J/\psi$ production, a
simplified method is used to calculate the amplitude square with
very good behavior in numerical calculations. But it cannot be applied to the calculation of $A$ and $\alpha$.

\begin{figure}
\center{
\includegraphics*[scale=0.38]{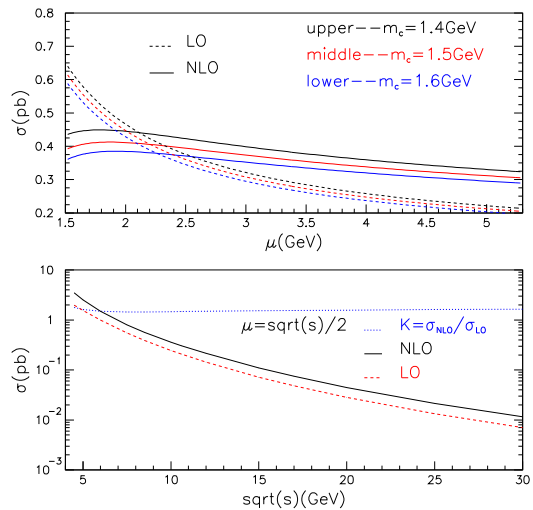}
\caption {\label{fig:scale}Cross sections as function of the renormalization scale $\mu$ and the center-of-mass energy
of $e^+e^-$ $\sqrt{s}$. }}
\end{figure}

\begin{figure}
\center{
\includegraphics*[scale=0.38]{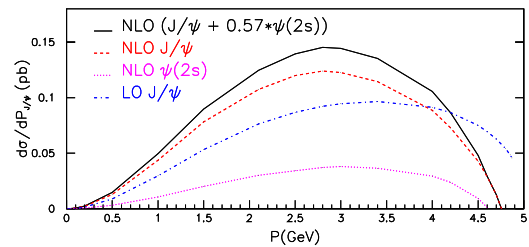}
\caption {\label{fig:dis}Momentum distribution of $\jpsi$ production with $m_c=1.5$ GeV and $\mu=2m_c$. 
}}
\end{figure}

\begin{figure}
\center{
\includegraphics*[scale=0.38]{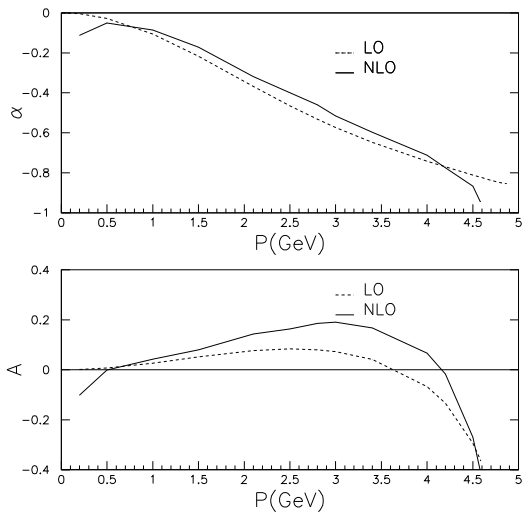}
\caption {\label{fig:polar}Polarization parameter $\alpha$ and angular distribution parameter $A$ of $\jpsi$ as functions of $p$ with $m_c=1.5$ GeV and $\mu=2m_c$.}}
\end{figure}

The values of $\a_s$ and the wave function at the origin of $\jpsi$
in the NLO calculation are taken the same as in
Ref.~\cite{Gong:2008ce}.
The numerical results are showed in Table.~\ref{table:result}.
\begin{table}[htbp]
\begin{center}
\begin{tabular}{|c|c|c|c|c|c|c|}
\hline\hline
$m_c$(GeV)&$\a_s(\mu)$&$\s^{(0)}$(pb)&$a(\hats)$&$\s^{(1)}$(pb)&$\s^{(1)}/\s^{(0)}$\\
\hline
1.4&0.267&0.341&2.35&0.409&1.20 \\
\hline
1.5&0.259&0.308&2.57&0.373&1.21\\
\hline
1.6&0.252&0.279&2.89&0.344&1.23\\
\hline\hline
\end{tabular}
\caption{Cross sections with different charm quark mass $m_c$ where the renormalization scale $\mu=2m_c$ and $\sqrt{s}=10.6 \gev$.}
\label{table:result}
\end{center}
\end{table}
The scale dependence of the cross section is shown at
Fig.~\ref{fig:scale} and it does improve significantly at NLO. The final
numerical result can be expressed as \be
\sigma^{(1)}=0.373^{+0.036}_{-0.079}\pb \label{eqn:result} \ee where
the theoretical uncertainty is from the choices of $m_c$ and
$\mu$, with $m_c=1.4\gev$ and $\mu=2m_c$ for the upper boundary 
and $m_c=1.6\gev$ and $\mu=\sqrt{s}/2$ for the lower boundary. The momentum distribution of $\jpsi$ production are shown in
Fig.~\ref{fig:dis}. To included the
$\psi^\prime$ contribution in the momentum distribution with a
suitable kinematic treatment, $m_c=m_{\psi^\prime}/2$,
$\mu=m_{\psi^\prime}$, $\mathrm{Br}(\psi^\prime \rightarrow J/\psi +
X)=0.574$ and $\Gamma(\psi^\prime\rightarrow ee)=2.19$ KeV are used.
In Fig.~\ref{fig:polar}, the momentum distributions of the
polarization factor $\a$ and the angular distribution coefficient
$A$ of $\jpsi$ are shown. Both $\alpha$ and $A$ have slight changes
at NLO. Furthermore, if the contribution from $\psi^\prime$ is included, the two curves change very little. In addition, we find that $d\s/d\mathrm{cos}\theta$ is a constant within quite large numerical error.

To further discuss the renormalization scale $\mu$ dependence, two ways are applied with $m_c=1.5$GeV. 
In one way, to find the maximum cross section by using ${{\partial
\sigma^{(1)}} / {\partial \mu}}=0, \sigma^{(1)}_{max}=0.41\pb$ is obtained
with $\mu=1.7$GeV, which means the calculation is right in the perturbative range and gives 
an uncertainty of about $0.04\pb$ for the default estimate. In another way, 
we apply the Brodsky, Lapage, Backenzie (BLM) scale setting\cite{Brodsky:1982gc} and find the unique scale choice $\mu^*=0.74$GeV with 
the cross section $\s^{(1)}(\mu^*)=\s^{(0)}(\mu^*)(1-9.1{\a_s(\mu^*)}/{\pi})$. It means that BLM way can 
not give good convergence behavior in this case. 

In summary, we have calculated the NLO QCD correction to
$e^+e^-\rightarrow \jpsi g g$ at the B factories. It increases the cross section to $0.373\pb$
with a K factor of about $1.21$ for the default $m_c=1.5\gev$ and $\mu=2m_c$.
By considering its dependence on the charm quark mass and
renormalization scale, the NLO cross section ranges from $0.294$ to
$0.409\pb$. Furthermore, it will be enhanced by another factor of about
$1.29$ and reaches $0.482\pb$ when the feed-down from $\psi^\prime$ is
considered.  By using  ${{\partial
\sigma^{NLO}(\mu) } / {\partial \mu}}=0$, a maximum $\sigma^{NLO}=0.41\pb$ is obtained
when $\mu=1.7$ GeV is chosen, which means the calculation is right in the
perturbative region. Together with the K factor $1.21$, it
could be expected that higher order corrections are smaller.
By considering theoretical and experimental uncertainty, our results
are roughly consistent with the measurement $0.6\pb$ from Belle~\cite{Abe:2001za,Abe:2002rb},
and furthermore agree well with their recent experimental measurement $0.43\pm 0.09 \pm 0.09\pb$~\cite{pakhlov:2008}. 
Thus there is little space left for the color-octet contribution now. 
There are large uncertainties in the calculation for the polarization via color-octet states when the color-octet states hadronize into color-singlet states. In contrast, it has much lesser uncertainties
in the calculation for the polarization of color-singlet state. In the photoproduction and hadronproduction of $\jpsi$, there are large discrepancies between the NLO theoretical predictions \cite{Gong:2008hk,Gong:2008ft} and experimental measurements for $\jpsi$ polarization. It may due to the large contribution from color-octet states with large uncertainties, or other mechanism \cite{Haberzettl:2007kj}. But in this process, color-singlet contribution is dominant and the convergence of perturbative QCD expansion works very well. Therefore, the prediction for its polarization distribution at NLO is well defined and should fit well with the experimental data. Our further work\cite{Gong:2009ng}
gives various distributions for $e^+e^-\rightarrow \jpsi+X(c\bar{c})$. 
In order to clarify the situation, we
suggest to perform further experimental analysis of the data based
on nowadays huge integrated luminosity at the  B factories. It is
desirable that the inclusive $\jpsi$ production could be separated
into $e^+e^-\rightarrow \jpsi+X(c\bar{c})$ and $e^+e^-\rightarrow
\jpsi+X(\mathrm{non}~c\bar{c})$ in experimental measurement, so that
their angular and momentum distributions of $J/\psi$
polarization can be compared with their theoretical predictions
separately. It may be worthwhile to include relativistic correction effects in order to sharpen the test of NRQCD. 

While this paper is being prepared, we are informed of the same process
also being considered by Ma, Zhang and Chao~\cite{Ma:2008gq}. Our results are in agreement with theirs. 
In addition, we calculated the momentum distribution of $J/\psi$ polarization which is a very 
important issue to clarify the $J/\psi$ polarization puzzle. 

We thank Y. Jia, J. W. Qiu, K. T. Chao and Y. Q. Ma for helpful comments and discussions. This work was supported by the National Natural Science Foundation of China (No.~10775141) and Chinese Academy of Sciences under Project No. KJCX3-SYW-N2.
\bibliography{paper}
\end{document}